\documentclass[letterpaper]{article} 
\usepackage{aaai25}  
\usepackage{times}  
\usepackage{helvet}  
\usepackage{courier}  
\usepackage[hyphens]{url}  
\usepackage{graphicx} 
\usepackage{natbib}  
\usepackage{caption} 
\usepackage{algorithm}
\usepackage{algorithmic}
\usepackage{amsmath}
\usepackage{amsthm}

\usepackage{newfloat}
\usepackage{listings}
\DeclareCaptionStyle{ruled}{labelfont=normalfont,labelsep=colon,strut=off} 
\floatstyle{ruled}
\newfloat{listing}{tb}{lst}{}
\floatname{listing}{Listing}
\title{Federated Learning and Free-riding in a Competitive Market}
\author{
    Jiajun Meng\textsuperscript{\rm 1}, Jing Chen\textsuperscript{\rm 2}, Dongfang Zhao\textsuperscript{\rm 3}, Lin Liu\textsuperscript{\rm 1*}
}
\affiliations{
    \textsuperscript{\rm 1}School of Economics and Management, Beihang University, Beijing 100191, China\\
    \textsuperscript{\rm 2}Department of Computer Science and Technology, Tsinghua University, Beijing 100084, China\\
    \textsuperscript{\rm 3}Tacoma School of Engineering and Technology, University of Washington, Seattle 98195, USA
    
    jjmeng@buaa.edu.cn, jchencs@mail.tsinghua.edu.cn, dzhao@cs.washington.edu, linliubh@buaa.edu.cn
}

\begin{document}

\maketitle

\begin{abstract}
Federated learning (FL) is a collaborative technique for training large-scale models while protecting user data privacy. Despite its substantial benefits, the free-riding behavior raises a major challenge for the formation of FL, especially in competitive markets, and hinders its applications and further development. In particular, to obtain a competitive advantage, a participating firm under FL has an incentive to take advantage of the global model without willingly contributing its information, discouraging information contribution by other firms and making FL formation collapse. Our paper explores this under-explored issue on how the free-riding behavior in a competitive market (in which both firms and consumers are strategic---firms are profit maximizing and consumers conduct self-selection of products) affects firms' incentives to form FL. Competing firms can improve technologies through forming FL to increase the performance of their products, which in turn, affects consumers' product selection and market size. The key complication is whether the free-riding behavior discourages information contribution by participating firms and results in the decomposition of FL, and even free riding does not discourage information contribution, this does not necessarily mean that a firm wants to form FL in a competitive market because free riding may reshape the competition positions of each participating firm and thus forming FL may not be profitable. We build a parsimonious game theoretical model that captures these interactions and our analyses show several new findings. First, even in the presence of the free-riding behavior, competing firms under FL find it optimal to contribute all its available information. Second, the firm with less amount of information always finds it profitable to free ride; whether its rival (with more amount of information) wants to form FL depends on the level of competition in the market and the gap in the amount of information possessed by participating firms; specifically, firms have an incentive to form FL when the level of competition or when the gap in information volume is not high. Third, when FL is formed, there exists an "All-Win" situation in which all stakeholders (participating firms, consumers, and social planner) benefit. Last, subsidizing by the free-riding firm can align its rival's incentive to form FL only when the level of competition is intermediate.
\end{abstract}

%

\section{Introduction}
Federated learning (FL) is an innovative collaborative technique that enables the training of large-scale models while preserving user data privacy. By distributing the training process across multiple entities, FL allows data to remain localized and secure, mitigating privacy concerns and adhering to data protection regulations \cite{kairouz2021}.

Despite its substantial benefits, the practical implementation of FL in competitive markets faces significant challenges. Among them, one major concern is the potential free-riding behavior, where competing firms may exploit the collective benefits of FL without actively contributing their information. This issue arises especially in a competitive market where strategic firms anticipate that sharing valuable information could inadvertently benefit their rivals \cite{Yang2019}. For example, in the industry of electric vehicles, firms strive to collect data related to driving, charging, as well as traffic and geographical information to consistently optimize autonomous driving and smart charging technologies. Undoubtedly, pooling information through a central server, FL collaboration can significantly improve the general performance of a participating firm's EV models. However, it may also strengthen the performance of its competitor's EV models, affecting the competitive position of each participating EV firm in the market. More importantly, this often constitutes a major concern that in a competitive market the free rider normally gains a competitive advantage and obtains a higher profit at the costs of other parties. Thus, in a competitive market, before deciding whether to participate in FL and contribute information, a firm has to weigh the potential negative force on its own profit brought by its rivals' free-riding behavior and the benefits associated with forming FL with them. Other industries like the ride-hailing market and financial sector also have the similar features and concerns. However, although there is a growing literature on FL formation with free-riding behavior \cite{CHEN2024120527,pmlr-v139-blum21a,9409833, Zeng2021ACS,zhu2021advanced,Lyu2020,9098045}, most of the studies have focused on a non-competitive environment which does not internalize the conflicts between the collective benefits of FL and the competitive incentives of participating firms. 

Our paper attempts to explore the under-explored issue on the interactions between FL formation and the free-riding behavior of firms in a competitive market. Specifically, we focus on the typical situation in which in the presence of the free-riding behavior FL can improve technologies that competing firms use to increase the performance of their products (e.g., in EVs industry and financial sector). We build an analytical model in which two competing firms decide whether to form FL, the optimal amount of information to contribute (if FL is formed), and their optimal prices by maximizing their respective expected profits, in anticipation of the expected improvement in performance as a result of FL and the potential free-riding behavior. The benchmark case is that each firm uses machine learning technology (ML) with its own data to improve the performance of its own product, and there is no free-riding behavior from its competing firms.

Thus, in a competitive market, the key trade-off faced by each firm when deciding whether to form FL is between obtaining more pooled information (than the amount of information it possesses) to better improve the performance of its own product and suffering from the free-riding behavior which helps its competing firm to obtain an improved competitive advantage. The key complication here is that when forming FL a firm might not want to contribute all its available information so as to mitigate the negative effect associated with the free-riding behavior from its competing firm. Collectively, if all participating firms under FL have the same incentive, then the pooled information under FL may not be necessarily greater in volume than that used under ML. If FL pools information at a low volume, then forming FL to improve the performance of products might result in an inferior performance level than that obtained under ML. And even if the amount of pooled information is greater under FL, this does not necessarily mean that a firm will yield a higher profit than that obtained under ML. Because its competing firm might obtain a competitive advantage through free-riding, which erodes its profit to a level lower than that obtained under ML.

This paper hopes to address the following research questions. First, in the presence of free-riding behavior, whether FL incentivizes the competing firms to use a subset of their information or contribute all the available information? And, is it possible that the free-riding concern let FL pool a lower amount of information than that used under ML? Second, how does the level of competition in the market affect firms' incentive in forming FL? Third, how does free-riding concern affect FL formation when firms have different amounts of information that can be contributed? Fourth, how does FL formation affect consumer surplus and social welfare, and whether there exists an "All-Win" situation in which all stakeholders benefit from FL formation? Last, under what condition can subsidizing from the free-riding firm encourage FL formation, and under what condition subsidizing fails?

Our paper shows several new findings. First, our results show that in a competitive market competing firms under FL find it optimal to contribute all its available information, even in the presence of the free-riding behavior. Thus, the free-riding behavior under FL does not discourage competing firms from contributing their information. This also implies that FL does pool a greater amount of information than that possessed by each participating firm, and thus, forming FL results in a higher level of product performance than that obtained under ML in which only the information of each individual firm is used. Second, although contributing all the information, the firm with less amount of information always finds it profitable to free ride under FL whereas its rival (with more information) wants to form FL only when the level of competition is low in the market or when the gap in information volume is limited; otherwise, the latter firm prefers to use ML to improve its product performance with its own data, and under this situation, FL formation collapses. Third, as long as FL is formed, there exists an ``All-Win'' situation in which all stakeholders (participating firms, consumers, and social planner) benefit, implying that the social planner should encourage the formation of FL. Last, to encourage the formation of FL, subsidizing by the free-riding firm can be considered. Our results show that this subsidizing can align its rival's incentive to form FL only when the level of competition is intermediate. Thus, the social planner should encourage such subsidizing by the free-riding firm for a higher social welfare and an "All-Win" situation, especially when the level of competition in the market is neither too low nor too high. We validate our key results through numerical experiments.

\section{Related Works}
\subsection{Federated Learning}
Federated learning (FL) is a decentralized machine learning paradigm that allows multiple entities to collaboratively train models without sharing their raw data. The process involves training local models on individual nodes and then aggregating the model updates on a central server \cite{pmlr-v54-mcmahan17a, konečný2017federatedlearningstrategiesimproving}. This decentralized approach ensures data privacy and security, as raw data never leaves its source. Many studies have focused on improving FL algorithms by optimizing communication efficiency, enhancing model aggregation techniques, and ensuring robust privacy-preserving mechanisms \cite{pmlr-v54-mcmahan17a, geyer2018differentiallyprivatefederatedlearning,MLSYS2019_7b770da6,NEURIPS2020_d37eb50d,9743558,zhang2024adversarial}. For example, the Federated Averaging (FedAvg) algorithm proposed by \citet{pmlr-v54-mcmahan17a} reduces communication costs and protects privacy by averaging model updates. \citet{zhang2024adversarial} proposes an adversarial attack framework to study the potential model-poisoned problem of FL.

To the best of our knowledge, our paper is the first study that considers FL formation in a competitive market in which strategic firms are competing for consumers whose demand depends on product features that are endogenously decided by firms. Specifically, the amount of information that each firm contributes under FL will affect the total amount of the pooled information, which in turn, influences the performance improvement of products; also, the selling prices are chosen by firms by maximizing their respective expected profits; product performance and prices both affect consumers' product selection and market size. Anticipating consumers' reactions, firms make their optimal decision on whether to form FL. This typical situation is widely seen in many industries with a competitive market in which consumers conduct self-selection of products. Our paper takes an initial step to explore such an under-explored issue and shows a series of results new to the literature of FL.

\subsection{Free Riding}
There is a growing body of research on the free-riding behavior in FL \cite{CHEN2024120527,doi:10.1287/mnsc.2023.00611,9705098,karimireddy2022mechanismsincentivizedatasharing, 9809786,9912903,karimireddy2022mechanismsincentivizedatasharing,pmlr-v139-blum21a,fraboni2021free,zhu2021advanced,Richardson2020,Lyu2020, 9098045}. Among these, \citet{Lyu2020} propose several fairness metrics for FL to reward high-contributing participants with greater returns. \citet{pmlr-v139-blum21a} introduce a framework for incentive-aware learning and data sharing to examine the incentive problem in participants’ contribution decisions. \citet{9705098} establish an infinitely repeated game to study the minimization of free-riders who do not participate in training the local models. \citet{doi:10.1287/mnsc.2023.00611} investigate the effect of contractual mechanisms on punishing free-riders and promoting sustained FL cooperation. 

However, these studies largely focus on a non-competitive situation, and thus, the formation of FL in these studies does not internalize the free-riding incentive of competing firms. To the best of our knowledge, this paper is the first study that explores the interactions between FL formation and the free-riding behavior in a competitive market. Our model captures the conflicts between the collective benefits of FL and the competitive incentives of participating firms caused by the free-riding behavior, and shows several new results on how competition and the free-riding behavior together affect the formation of FL.

\section{Model Setup}
Consider a competitive market with two firms (firms 1 and 2), each selling a product. Let $p_1$ and $p_2$ denote the prices chosen by the two firms and $q_1$ and $q_2$ the demands of the two products. Following the classical framework \cite{Dixit1979, Vives1984}, firms' prices and demands have the following (negative) relationship that is given by

\begin{equation}\label{vives}
\left\{
\begin{aligned}
p_1&= v_1-q_1-\gamma q_2\\
p_2&= v_2-q_2-\gamma q_1,\\
\end{aligned}
\right.
\end{equation}
where $v_i$ represent the quality of product $i$ consumed ($i=1$ or $2$) and \( \gamma \) represents the degree of product substitution ($\gamma \in [0,1]$). Specifically, a larger value of \( \gamma \) means that both products are closer substitutes; when $\gamma=1$ ($\gamma=0$), both products are perfect substitutes (both products do not substitute each other at all). Without loss of generality, we let product 1 have a (weakly) higher quality (i.e., \( v_1 \ge v_2 > 0 \)). Thus, firm 1 (firm 2) is the dominant firm (weaker one). For simplicity, we ignore the production costs of both firms.

Suppose that each firm has a dataset which includes $\mathcal{D}_i$ amount of information that can be used to contribute to form FL. We consider two situations: (1). the dominant firm 1 has more data (i.e., $\mathcal{D}_1>\mathcal{D}_2\geq 0$) and (2). the weaker firm 2 has (weakly) more data (i.e., $\mathcal{D}_2\ge\mathcal{D}_1\geq 0$).

Each firm has two strategies that can be used to improve the quality of its product ($v_i$): (1) using ML training independently with its own data or (2) forming FL with pooling information from both firms. For both strategies, we use $f(\cdot)$ to represent the training effectiveness in improving the product quality as a function of the total amount of information used in training $\mathcal{R}$, where $f(\cdot)$ is concave in $ \mathcal{R}$ (i.e., $f'(\mathcal{R}) > 0$  and  $f''(\mathcal{R}) < 0$), implying that as the amount of information used increases the training effectiveness increases at a decreasing rate. Let $\mathcal{R}_i$ represent the amount of information used by firm $i$. Thus, when firms use ML training, the updated level of quality for firm $i$ is given by $v_i+f(\mathcal{R}_i)$, which is independent of the rival's information used ($\mathcal{R}_{-i}$).

Alternatively, when using FL formation, each firm contributes $\mathcal{R}_i$ amounts of information for local training. the generated training outputs (e.g., gradients) are uploaded to a central server and aggregated through FL algorithms (e.g., FedAvg) to build a global model that can be used by both firms. For simplicity, we assume that (1) there is no information overlap of the information possessed by both firms (i.e., the amount of pooled information is given by $\mathcal{R}_i+\mathcal{R}_{-i}$) and (2) the training effectiveness under FL is the same for both firms ($f(\mathcal{R}_i+\mathcal{R}_{-i})$). Because the training outcome of FL depends on the amount of pooled information \cite{Sheller2020FederatedLI}, the updated level of quality for firm $i$ under FL is given by $v_i+f(\mathcal{R}_i+\mathcal{R}_{-i})$. 

It is worth noting that this above formulation implies that to gain a competitive advantage firm $i$ might have an incentive to free ride its rival by contributing less amount of information. Such a free-riding behavior, however, does not emerge under ML because the improved quality of one firm does not depend on the amount of used information by the other.

In addition, we assume that the quality improvement (under both strategies) does not alter the degree of product substitution (i.e., $\gamma$ is not affected by firms' strategy). In the following text, we use superscript $ml$ ($fl$) to represent the case with ML training (FL formation). 

Specifically, when firms use ML training, the inverse demand functions in (\ref{vives}) become:

\begin{equation}\label{mlp}
\left\{
\begin{aligned}
p_1&= v_1^{ml}(\mathcal{R}_1)-q_1-\gamma q_2\\
p_2&= v_2^{ml}(\mathcal{R}_2)-q_2-\gamma q_1, \\
\end{aligned}
\right.
\end{equation}
where $v_i^{ml}(\mathcal{R}_i)$ is the updated quality of product $i$ under ML and is equal to $v_i+f(\mathcal{R}_i)$ and the second term $f(\mathcal{R}_i)$ represents the improved level of quality of product $i$ under ML with firm $i$'s own data $\mathcal{R}_i$. Alternatively, when firms use FL collaboration, the inverse demand functions in (\ref{vives}) become:

\begin{equation}\label{flp}
\left\{
\begin{aligned}
p_1&= v_1^{fl}(\mathcal{R}_1,\mathcal{R}_2)-q_1-\gamma q_2\\
p_2&= v_2^{fl}(\mathcal{R}_1,\mathcal{R}_2)-q_2-\gamma q_1,\\
\end{aligned}
\right.
\end{equation}
where $v_i^{fl}(\mathcal{R}_i,\mathcal{R}_{-i})$ is the updated quality of product $i$ under FL and is equal to $v_i+f(\mathcal{R}_i+\mathcal{R}_{-i})$ and the second term $f(\mathcal{R}_i+\mathcal{R}_{-i})$ represents the improved level of quality for both products under FL collaboration. Assume that firms are rational and choose the strategy that yields a higher (positive) profit. Notably, if at least one firm chooses not to form FL, then the situation degenerates to that in which each firm uses separated ML training with its own data. That is, FL is formed only when both firms find it profitable to do so.

The timing of the game is as follows: First, each firm decides whether to form FL: if both firms opt for FL, they then decide how much information $\mathcal{R}_i$ to contribute, where $0<\mathcal{R}_i\leq\mathcal{D}_i$; otherwise, each firm uses ML training with its own data. Second, each firm simultaneously decides its optimal price. Last, demands are realized and profits are obtained by each firm. We use backward induction to solve the game.

\begin{figure}[htbp]
    \centering
    \includegraphics[width=1\linewidth]{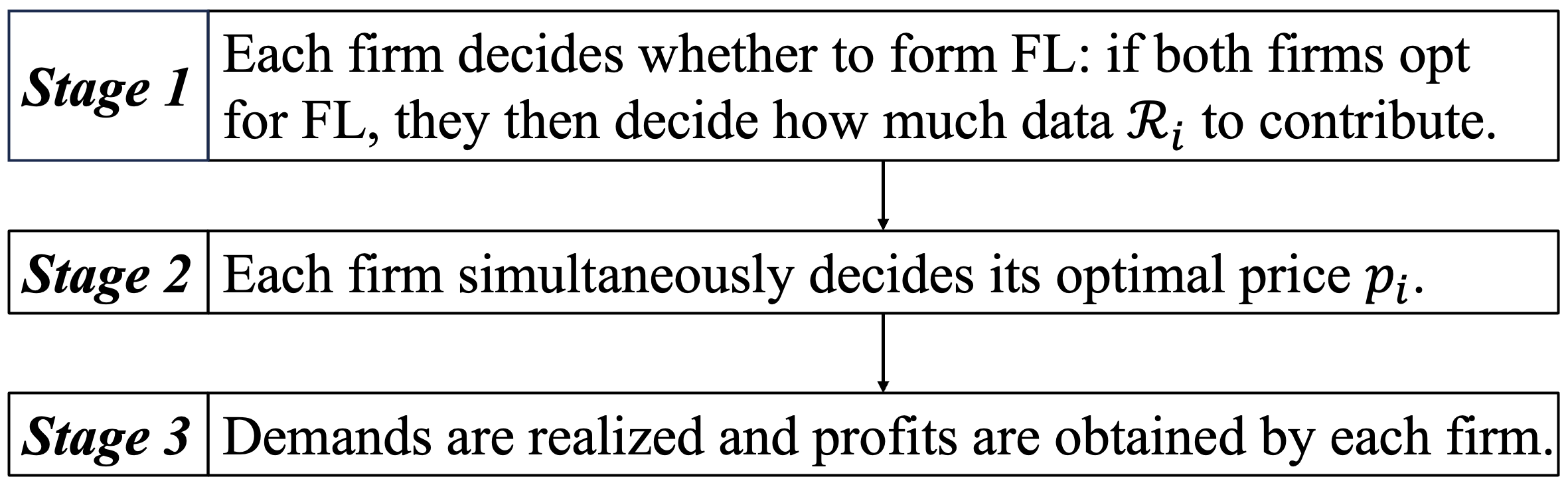}
    \caption{Three-stage game.}
    \label{fig1}
\end{figure}

To avoid the trivial cases in which one firm is competed out of the market (i.e., to ensure a competitive market), the following condition is required:

\subsubsection{Condition 1.} \textit{Product quality ratio should be sufficiently low, (i) $1 \!\le\! \frac{v_1}{v_2}\!<\!\frac{2-\gamma^2}{\gamma}$ and (ii) $\frac{\gamma}{2-\gamma^2}\!<\!\frac{v_{-i}+f(\mathcal{R}_{-i})}{v_i+f(\mathcal{R}_{i})}\!<\!\frac{2-\gamma^2}{\gamma}$}.
\\ \hspace*{\fill} \\
\indent Condition (i) ensures a competitive market when firms do not implement ML or FL. Specifically, the quality difference between the two firms is not too large so that the dominant firm 1 does not have a strong competitive advantage over the weaker firm 2 which forces it out of the market. Condition (ii) guarantees a competitive market when firms use ML: notably, because under ML it is possible that the weaker firm 2 eventually has a stronger updated quality (due to a potentially greater amount of training data used) that drives the dominant firm 1 out of the market, this condition applies to both firms and thus uses $i$ and $-i$. It is worth noting that condition (i) is sufficient to ensure a competitive market when firms form FL because under FL both firms have the same improved quality level ($f(\mathcal{R}_i+\mathcal{R}_{-i})$), implying that the dominant firm 1 still has a higher updated quality level under FL and thus the condition that is required to ensure a competitive market under FL ($1 \le \frac{v_1+f(\mathcal{R}_1+\mathcal{R}_2)}{v_2+f(\mathcal{R}_1+\mathcal{R}_2)}<\frac{2-\gamma^2}{\gamma}$) holds naturally.

\section{Equilibrium Analysis}
In this section, we will explore firms' two strategies used to improve the quality of their products: (1) each firm separately uses ML training with its own data and (2) both firms use FL collaboration with pooled information from both firms. The first strategy with ML training is used as the benchmark, and a firm will consider to form FL only when it finds doing so yields a higher expected profit than that obtained under ML training. Proofs of all lemmas and theorems can be found in the appendix.

\subsection{Machine Learning (ML)}
Rearranging the equations in (\ref{mlp}), one can easily confirm that the (direct) demand of product $i$ is given by 

\begin{equation}\label{mlq}
q_i(p_i,p_{-i})\!=\!\frac{v_i^{ml}(\mathcal{R}_i)\!-\!p_i\!-\!\gamma  [v_{-i}^{ml}(\mathcal{R}_{-i})-\!p_{-i}]}{1-\gamma ^2},
\end{equation}
where $q_i(p_i,p_{-i})$ increases in the updated quality of product $i$ ($v_i^{ml}(\mathcal{R}_i)$) and the price of competing product ($p_{-i}$) while decreases in the updated quality of the competing product ($v_{-i}^{ml}(\mathcal{R}_{-i})$) and its own price ($p_i$). The (expected) profit of firm $i$ is given by $\Pi_i^{ml}(p_i,p_{-i})=p_i q_i(p_i,p_{-i})$, where $p_i$ is given by (\ref{mlp}) and $q_i(p_i,p_{-i})$ is given by (\ref{mlq}). 

We use the backward induction to solve the game. Given the amount of data used under ML ($\mathcal{R}_1, \mathcal{R}_2$), firm $i$ chooses its optimal \(p_i\) by maximizing its expected profit. Then, firm $i$ chooses the optimal amount of data used under ML subject to $0<\mathcal{R}_i^{ml} \le \mathcal{D}_i$. The following lemma summarizes the results.

\subsubsection{Lemma 1.} \textit{In equilibrium, each firm uses all its data (i.e., $\mathcal{R}_i^{ml} =\mathcal{D}_i$) and the optimal price is given by $p^{ml}_i\!=\!\frac{  \left(2\!-\!\gamma ^2\right) v_i^{ml}(\mathcal{D}_i)\!-\!\gamma  v_{-i}^{ml}(\mathcal{D}_{-i})}{4-\gamma ^2} $; firm $i$'s profit is given by $\Pi^{ml}_i\!=\!\frac{\left[  \left(2\!-\!\gamma ^2\right) v_i^{ml}(\mathcal{D}_i)\!-\!\gamma  v_{-i}^{ml}(\mathcal{D}_{-i})\right]^2}{(1-\gamma^2)(4-\gamma ^2)^2}$, where $v_i^{ml}(\mathcal{D}_i)=v_i+f(\mathcal{D}_i)$}.
\\ \hspace*{\fill} \\
\indent This lemma illustrates that in equilibrium each firm will use all its data to improve the quality of its product to the highest possible level. Specifically, using more data brings two benefits: first, it increases a firm's pricing power in the competitive market and thus allows it to charge a higher price; second, a higher quality level expands the market and thus lets the firm obtain a greater demand. These two positive forces together imply that using more data yields a higher profit. Thus, each firm wants to use all its data to improve its product quality as much as possible under ML.

\subsection{Federated Learning (FL)}
Next, we explore firms' FL formation in a competitive market. Rearranging the equations in (\ref{flp}), one can easily confirm that the (direct) demand of product $i$ is given by 

\begin{equation}\label{flq}
q_i(p_i,p_{-i})\!=\!\frac{v_i^{fl}(\mathcal{R}_i,\mathcal{R}_{-i})\!-\!p_i\!-\!\gamma  [v_{-i}^{fl}(\mathcal{R}_i,\mathcal{R}_{-i})-\!p_{-i}]}{1-\gamma ^2},
\end{equation}
where $q_i(p_i,p_{-i})$ increases in the updated quality of product $i$ ($v_i^{fl}(\mathcal{R}_i,\mathcal{R}_{-i})$) and the price of competing product ($p_{-i}$) while decreases in the updated quality of the competing product ($v_{-i}^{fl}(\mathcal{R}_i,\mathcal{R}_{-i})$) and its own price ($p_i$). The profit of firm $i$ is given by $\Pi_i^{fl}(p_i,p_{-i})=p_i q_i(p_i,p_{-i})$ where $p_i$ is given by (\ref{flp}) and $q_i(p_i,p_{-i})$ is given by (\ref{flq}). 

Similarly, the backward induction is used to solve the game. Given the amount of information used under FL ($\mathcal{R}_1, \mathcal{R}_2$), firm $i$ chooses its optimal \(p_i\) by maximizing its expected profit. Then, firm $i$ chooses the optimal amount of information used under FL subject to $0<\mathcal{R}_i^{fl} \le \mathcal{D}_i$. The following lemma summarizes the results.

\subsubsection{Lemma 2.} \textit{In equilibrium, each firm will share all its information (i.e., $\mathcal{R}_i^{fl} =\mathcal{D}_i$) and the optimal price is given by $p^{fl}_i\!=\!\frac{  \left(2\!-\!\gamma ^2\right) v_i^{fl}(\mathcal{D}_i,\mathcal{D}_{-i})\!-\!\gamma  v_{-i}^{fl}(\mathcal{D}_i,\mathcal{D}_{-i})}{4-\gamma ^2} $; firm $i$'s profit is given by $\Pi^{fl}_i\!=\!\frac{\left[  \left(2\!-\!\gamma ^2\right) v_i^{fl}(\mathcal{D}_i,\mathcal{D}_{-i})\!-\!\gamma  v_{-i}^{fl}(\mathcal{D}_i,\mathcal{D}_{-i})\right]^2}{(1-\gamma^2)(4-\gamma ^2)^2}$}, where $v_i^{fl}(\mathcal{D}_i,\mathcal{D}_{-i})=v_i+f(\mathcal{D}_i+\mathcal{D}_{-i})$.
\\ \hspace*{\fill} \\
\indent The key message delivered in this lemma is that even under FL (where free-riding behavior emerges) each firm still wants to share all its information ($\mathcal{R}_i^{fl} =\mathcal{D}_i$). Specifically, compared to ML situation, sharing a greater amount of information under FL creates an additional force--letting its rival to free ride the quality improvement. However, this negative force is dominated by the two above-mentioned positive forces associated with price and demand, implying that under FL sharing a greater amount of its information yields a higher profit. As a result, in equilibrium, each firm will contribute all its available information under FL formation.

Combining the results of Lemmas 1 and 2, one can conclude that under both ML and FL each firm wants to use all its available information. However, although the free-riding behavior in the competitive market does not alter firms' incentive to use all their available information, it might affect their choices on which specific strategy (ML or FL) to use to improve the quality of their products. This might depend on parameters for the level of competitiveness of the market (e.g., the degree of product substitution $\gamma$) and the relative competition positions of both firms (e.g., product qualities $v_i$ and the amount of available information $\mathcal{D}_i$).

It is worth noting that the results in Lemma 2 that both firms use all their available information also imply that the firm with less amount of information always has an incentive to free ride under FL in the competitive market. Thus, in the following analyses, when exploring firms' optimal strategy (ML vs. FL), we discuss two situations (1) firm 1 has more information (i.e., $\mathcal{D}_1>\mathcal{D}_2\geq 0$) and (2) firm 2 has (weakly) more information (i.e., $\mathcal{D}_2\ge\mathcal{D}_1\geq 0$).

In the following main text, we will focus on the situation in which firm 1 has more information (i.e., $\mathcal{D}_1>\mathcal{D}_2\geq 0$). The section with additional analyses includes the other situation in which firm 2 has (weakly) more information (i.e., $\mathcal{D}_2\ge\mathcal{D}_1\geq 0$).

\subsection{Strategy Selection: ML or FL}
Each firm will choose the strategy (ML or FL) that provides the higher (expected) profit. Specifically, FL is formed only when each firm obtains a higher profit under FL (i.e., $\Pi^{fl}_1>\Pi^{ml}_1$ and $\Pi^{fl}_2>\Pi^{ml}_2$ hold at the same time). Otherwise, both firms use ML strategy.

When firm 1 has more information ($\mathcal{D}_1>\mathcal{D}_2\geq 0$), firm 2 has an incentive to free ride under FL. Thus, the key trade-off faced by firm 1 is between (i) using its own data to improve the quality of its product independently and thus avoiding firm 2's free-riding behavior (ML strategy), and (ii) pooling information with firm 2 but suffering from its free-riding behavior (FL strategy). This trade-off of firm 1 is illustrated in the following lemma.

\subsubsection{Lemma 3.} \textit{ $\Pi_2^{fl}>\Pi_2^{ml}$   always holds, while  $\Pi_1^{fl}>\Pi_1^{ml}$ does not necessarily hold.}
\\ \hspace*{\fill} \\
\indent This lemma illustrates that firm 2 always has an incentive to form FL because its free-riding behavior under FL yields a higher profit than that obtained under ML ($\Pi_2^{fl}>\Pi_2^{ml}$). However, firm 1 might be reluctant to form FL because it does not want to help firm 2 to improve its competitive position through its improved product quality obtained under FL ($\Pi_1^{fl}>\Pi_1^{ml}$ may not hold). The following theorem summarizes the condition that ensures that firm 1 also finds it profitable to form FL ($\Pi_1^{fl}>\Pi_1^{ml}$ holds too).

\subsubsection{Theorem 1.} \textit{FL is formed if and only if condition (\ref{Thm1}) is satisfied}
\begin{equation}\label{Thm1}
1< \frac{f(\mathcal{D}_1+\mathcal{D}_2)-f(\mathcal{D}_2)}{f(\mathcal{D}_1+\mathcal{D}_2)-f(\mathcal{D}_1)}<\frac{2-\gamma^2}{\gamma}.
\end{equation}
\\ \hspace*{\fill} \\
\indent This theorem illustrates the condition that is required to incentivize firm 1 to form FL (recall from Lemma 3 that firm 2 always wants to form FL). Specifically, under both strategies, each firm wants to use all its information (see Lemmas 1 and 2), implying that for each firm the improved quality under FL is higher than that under ML ($f(\mathcal{D}_1+\mathcal{D}_2)>f(\mathcal{D}_i)$). Thus, condition (\ref{Thm1}) ensures that the net improved quality (from ML to FL) for firm 2 (i.e., $f(\mathcal{D}_1+\mathcal{D}_2)-f(\mathcal{D}_2)$) is not too large compared to that for firm 1 (i.e., $f(\mathcal{D}_1+\mathcal{D}_2)-f(\mathcal{D}_1)$). Under this condition, the net benefit of the free-riding behavior is limited for firm 2, implying that firm 1 is still able to keep its competitive advantage over firm 2 at a sufficiently high level under FL. And, the pooling information through FL can significantly improve the net benefit for firm 1 because $f(\mathcal{D}_1+\mathcal{D}_2)-f(\mathcal{D}_1)$ is sufficiently large, implying an increased pricing power and expanded demand for firm 1 (compared to those obtained under ML). As a result, firm 1 finds it optimal to form FL with firm 2, even in the presence of the free-riding behavior.

The next result discusses how parameters for the level of competitiveness of the market (the degree of product substitution $\gamma$) and the relative competition positions of both firms (product qualities $v_i$ and the amount of available information $\mathcal{D}_i$) affect the formation of FL. The following theorem summarizes the results.

\subsubsection{Theorem  2.} 
\textit{(i) There exists a threshold degree of product substitution (\( \gamma^* \in (0,1)\)) such that FL is formed if \( 0 < \gamma < \gamma^* \); otherwise, ML is formed.} 

\textit{(ii) There exists a threshold amount of information for firm 2 (\( \mathcal{D}_2^* \)) such that FL is formed if \( \mathcal{D}_2^* < \mathcal{D}_2< \mathcal{D}_1 \); otherwise, ML is formed.}

\textit{(iii) The (initial) product quality (\( v_i \)) does not affect firms' preference for ML and FL.}
\\ \hspace*{\fill} \\
\indent  This theorem delivers several key messages. First, the degree of product substitution ($\gamma$) affects firms' incentive on which strategy to use (FL or ML). Specifically, when the degree of product substitution is low (\( 0 < \gamma < \gamma^* \)), FL is formed (otherwise, ML is formed). In particular, in the former situation with a low degree of product substitution, firms' competition is not high because one product cannot be easily substituted by the other. Thus, under such a market with a low level of competition, the negative effect associated with the free-riding behavior by firm 2 under FL that reduces firm 1's profit is weak, and thus, is dominated by the two above-mentioned positive effects associated with using the pooled information to increase the product quality so that both pricing power and the market size can be higher. Thus, FL is formed. However, in the latter situation with a high degree of product substitution, as products can be easily substituted by each other, firms are competing more aggressively in the market. Under this situation, the negative effect associated with the free-riding behavior by firm 2 constitutes the major concern for firm 1, and thus, dominates the two positive effects of using pooled information to increase pricing power and market size. As a result, firm 1 finds it more profitable to use ML and use its own data to train product quality rather than form FL with firm 2. This explains part (i).

The second result is about how the amount of available information ($D_i$) affects firms' preference on FL and ML: when the free-riding firm (firm 2) has a large amount of information (\( \mathcal{D}_2^* < \mathcal{D}_2< \mathcal{D}_1 \)) FL is formed (otherwise, ML is formed). Specifically, when firm 2 has a large amount of information (but still less than that of firm 1), the difference in the amount of information contributed by both firms under FL is small (recall that each firm uses all its available information under FL, see Lemma 2), implying that the negative effect associated with the free-riding behavior is small, and thus, is dominated by the two positive effects associated with increasing prices and market size. Thus, firm 1 finds it optimal to form FL with firm 2. In contrast, when firm 2 has a small amount of information, the negative effect associated with free-riding is strong and dominates the two positive effects, implying that firm 1 is reluctant to form FL with firm 2 and thus uses ML instead. This explains part (ii).

Third, firms' preference for FL and ML is not affected by their (initial) product qualities (\( v_i \)). This is because when selecting between FL and ML firm 1 compares the profit obtained under both strategies associated with the improved level of quality (associated with the term $f(\cdot)$ in (\ref{mlp}) and ({\ref{flp}})), and thus, the (initial) product qualities (\( v_i \)) do not play a role in affecting firms' preferences over FL and ML. This explains part (iii).

\subsection{Welfare Analyses} 
The above analyses focus on firms' incentives and strategy selection. The next result explores how FL and ML affect consumer surplus and social welfare. We use CS and SW to respectively denote consumer surplus and social welfare. The following theorem summarizes the results.

\subsubsection{Theorem 3.} \textit{When FL is formed (condition (6) satisfied), its consumer surplus is always higher ($CS^{fl}>CS^{ml}$) and there is an “All-Win” situation in which all stakeholders (firms, consumers, and social planner) benefit from forming FL.}
\\ \hspace*{\fill} \\
\indent This theorem illustrates that as long as firm 1 finds it optimal to form FL with firm 2 (condition (\ref{Thm1}) is satisfied), consumer surplus is higher than that under ML because product qualities are improved with pooled information from both firms, and thus, are higher than those under ML. This directly increases consumer surplus under FL through a higher willingness to pay by consumers. In addition, the second positive effect on consumer surplus is that the market size is expanded because of the higher qualities. Although this also incentivizes firms to charge higher prices, which reduces consumer surplus, this negative effect is dominated by the two positive effects, implying that consumer surplus is higher under FL.

It is worth noting that social welfare is also higher under FL as long as firm 1 wants to form it with firm 2. This is because price is merely a surplus transfer between consumers and firms and thus only the benefits to consumers associated with higher willingness to pay and expanded market play a role here, implying that social welfare is higher under FL. Obviously, these results imply that when FL is formed there is an “All-Win” situation in which all stakeholders (firms, consumers, and social planner) benefit.

\begin{figure*}[htbp]
    \centering
    \includegraphics[width=0.92\linewidth]{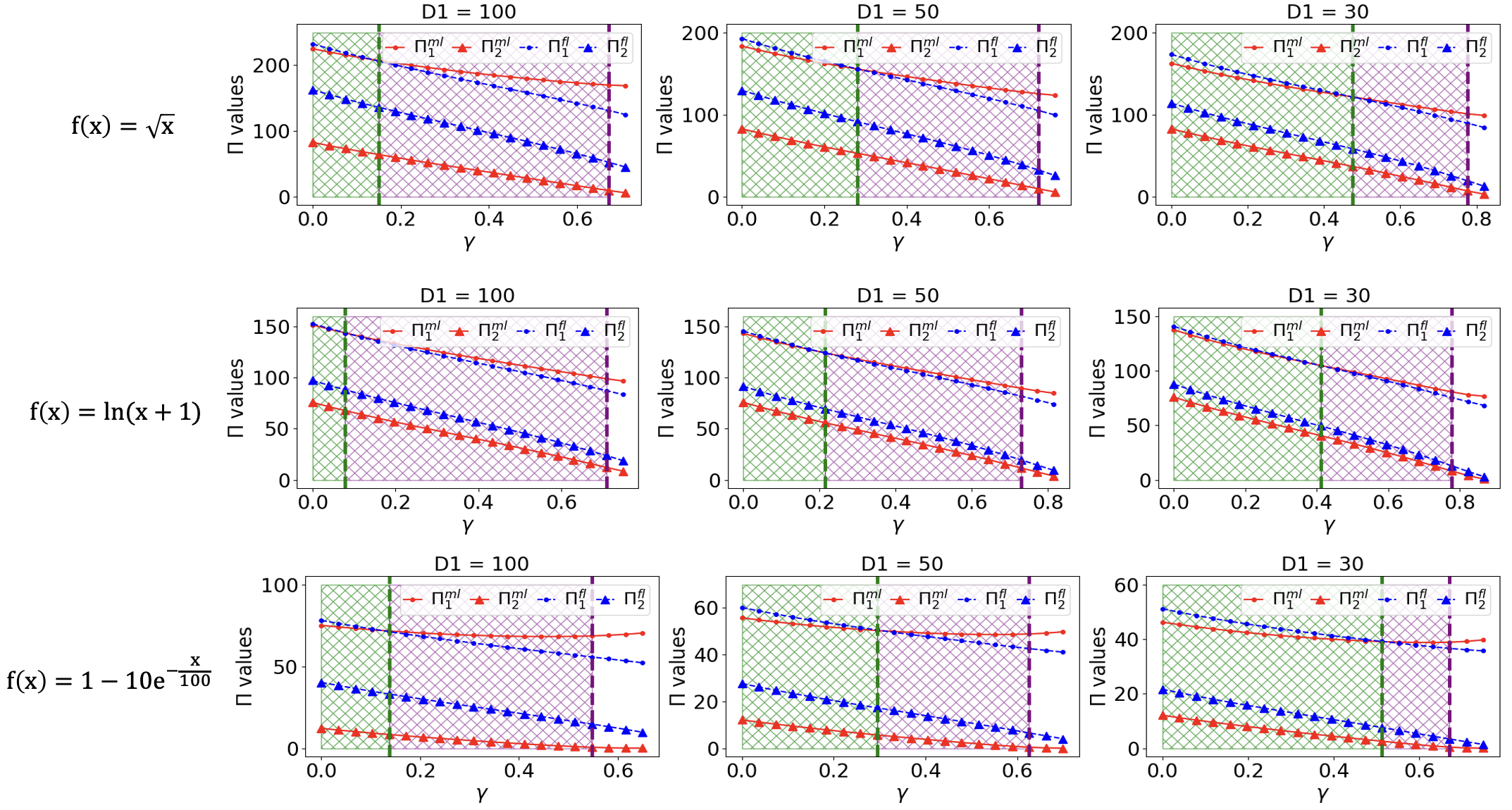}
    \caption{Firms' Equilibrium Profits of FL and ML.}
\end{figure*}

\section{Additional Analyses}
Next, we conduct some additional analyses, including the situation in which firm 2 has more information, subsidizing firm 1 to form FL with firm 2, and numerical experiments.

\subsection{Firm 2 Has More Information: $\mathcal{D}_2\ge\mathcal{D}_1\geq 0$}
Recall that firms 1 and 2 have different (initial) product qualities (\( v_1 \ge v_2>0 \)) and  Theorem 2 (iii) indicates that the (initial) product quality (\( v_i \)) does not affect firms' preference for ML and FL (condition (\ref{Thm1}) does not include (\( v_i \))). In addition, all other conditions (except (i) in Condition 1 which ensures a competitive market before neither ML nor FL is implemented) also do not include (\( v_i \))). Thus, one can easily conclude that when firm 2 has more information firm 1 has an incentive to free ride and the rest of the results also mirror those obtained in the main analyses (with firm 1 and firm 2 exchanged in conditions and discussions).

\subsection{Subsidizing}
In the main analyses, firm 1 might be reluctant to form FL with firm 2 because its free-riding behavior might harm firm 1's profit in the competitive market (see Lemma 3). In this section, we explore the situation in which the free-riding firm 2 might subsidize firm 1 to incentivize FL formation.

According to Lemma 3 and part (i) in Theorem 2, firm 2 always finds it profitable to form FL but firm 1 may not, and firm 1 wants to form FL with firm 2 only when the degree of product substitution is low (otherwise, ML is used). Thus, to incentivize firm 1 to form FL, firm 2 may want to subsidize firm 1. Specifically, the profit improvement for firm 2 under FL is given by $\Delta_2=\Pi_2^{fl}-\Pi_2^{ml}$ while the profit loss for firm 1 is given by $\Delta_1=\Pi_1^{fl}-\Pi_1^{ml}$. Thus, as long as $\Delta_2\geq \lvert\Delta_1 \rvert$ (or equivalently $\Delta_1+\Delta_2 \ge 0$) is satisfied, subsidizing might align both firms' incentives in FL formation. The following theorem summarizes the results.

\subsubsection{Theorem 4.} \textit{There exists a threshold degree of product substitution ($\hat{\gamma} \in (\gamma^*,1)$) such that subsidizing works for FL formation if \(\gamma^* < \gamma < \hat{\gamma}\); otherwise, subsidizing fails.}
\\ \hspace*{\fill} \\
\indent This theorem illustrates that subsidizing by firm 2 can incentivize firm 1 to form FL only when the degree of product substitution is intermediate (i.e., \(\gamma^* < \gamma < \hat{\gamma}\)). Specifically, for a lower degree of substitution (\( 0 \le \gamma < \gamma^* \)), without any subsidy from firm 2, firm 1 wants to form FL (see part (i) of Theorem 2) because the market is not very competitive and thus the free-riding behavior of firm 2 just slightly hurts the profit of firm 1 but is eventually dominated by the benefits associated with FL. However, for a very high degree of substitution ($\hat{\gamma}\ < \gamma \le 1$), the market is so competitive that the free-riding behavior of firm 2 reduces the profit of firm 1 badly, and thus, the profit improvement of firm 2 cannot compensate for the profit loss of firm 1. As a result, subsidizing by firm 2 cannot incentivize firm 1 to form FL.

\subsection{Numerical Experiments}
We conduct numerical analyses to illustrate the previous main results. Specifically, three functions are used to capture $f(\cdot)$: $f(x)=\sqrt{x}$, $f(x)=\ln{(x+1)}$, and $f(x)=1-10 e^{-\frac{x}{100}}$. The parameters are set as follows: $v_1=20$, $v_2=15$, $\mathcal{D}_1\in\{100, 50, 30\}$ and $\mathcal{D}_2=10$. In Figure 3, the x-axes represents the degree of product substitution ($\gamma$) and the y-axes represents firms' profits under ML and FL, and we plot figures with different $\mathcal{D}_1$ values for each function.

From Figure 3, one can easily confirm that Lemma 3 holds because in each subfigure the thick blue line is always above the think red line ($\Pi_2^{fl}>\Pi_2^{ml}$) whereas the thin blue line is not ($\Pi_1^{fl}>\Pi_1^{ml}$ does not necessarily hold). In addition, the green vertical dashed line represents $\gamma=\gamma^*$, i.e., the case where $\Pi_1^{ml}=\Pi_1^{fl}$, and thus, the region in green shows the parameter space in which firm 1 wants to form FL with firm 2 (see Theorem 2). The purple vertical dashed line represents $\gamma=\hat{\gamma}$, i.e., the case where $\Delta_1+\Delta_2 =0$, and thus, the region in purple shows the parameter space in which subsidizing by firm 2 can incentivize firm 1 to form FL, whereas subsidizing fails in the region in white in which the market is so competitive that the free-riding behavior of firm 2 hurts the profit of firm 1 so much and thus subsidizing by firm 2 cannot incentivize firm 1 to form FL any more (see Theorem 4). Also, these figures show that as $\mathcal{D}_1$ decreases (the difference between the amount of information available for each firm is smaller), the region in green expands, implying firm 1 wants to form FL with firm 2 in a greater area (see Theorem 2). These results hold for all the three functions for $f(\cdot)$.

\section{Conclusion}
We develop a game theoretical model to explore FL formation in a competitive market where the free-riding behavior is the key concern. In particular, we focus on the typical situation in which FL can improve technologies used by competing firms to increase the performance of their products (e.g., in EVs industry and financial sector). Our model captures their competitive incentives in the market as well as their collective incentives in FL formation in the presence of the free-riding behavior. Our analyses show a series of interesting results which are validated by numerical experiments.

\section{Appendix}

\subsection{Proof of Lemma 1} 

The (expected) profit of firm $i$ is $\Pi_i^{ml}(p_i,p_{-i})=p_i\!\frac{v_i^{ml}(\mathcal{R}_i)\!-\!p_i\!-\!\gamma  [v_{-i}^{ml}(\mathcal{R}_{-i})-\!p_{-i}]}{1-\gamma ^2}$. By  solving $\frac{\partial \Pi_1^{ml}(p_1,p_2)}{\partial p_1}\!=\!0$ and $\frac{\partial \Pi_2^{ml}(p_1,p_2)}{\partial p_2}\!=\!0$, we obtain firm $i$’ optimal prices (given the amount of data used $\mathcal{R}_1$ and $\mathcal{R}_2$) given by $p^{ml}_i(\mathcal{R}_i,\mathcal{R}_{-i})\!\!=\!\!\frac{  \left(2\!-\!\gamma ^2\right) v_i^{ml}(\mathcal{R}_i)\!-\!\gamma  v_{-i}^{ml}(\mathcal{R}_{-i})}{4-\gamma ^2}$. One can easily verify that the second-order condition holds and thus $p^{ml}_i(\mathcal{R}_i,\mathcal{R}_{-i})$ constitutes a local maximizer.

Substituting $p^{ml}_i(\mathcal{R}_i,\mathcal{R}_{-i})$ back to $\Pi_i^{ml}(p_i,p_{-i})$ yields $\Pi_i^{ml}(\mathcal{R}_i,\mathcal{R}_{-i})\!=\!\frac{\!\left[\!  \left(2\!-\!\gamma ^2\right) v_i^{ml}(\mathcal{R}_i)\!-\!\gamma  v_{-i}^{ml}(\mathcal{R}_{-i})\!\right]\!^2}{(1-\gamma^2)(4-\gamma ^2)^2}$. Firm $i$ chooses the optimal amount of data used for training. One can easily verify that $\frac{\partial \Pi_i^{ml}(\mathcal{R}_i,\mathcal{R}_{-i})}{\partial \mathcal{R}_i}\!=\!\frac{2(2-\gamma^2)\left[  \left(2-\gamma ^2\right) v_i^{ml}(\mathcal{R}_i)-\gamma  v_{-i}^{ml}(\mathcal{R}_{-i})\right]}{(1-\gamma^2)(4-\gamma ^2)^2}\cdot \frac{\partial f(\mathcal{R}_i)}{\partial \mathcal{R}_i}>0$ holds, because of (ii) in Condition 1, $0\le \gamma\le 1$, and $f'(\mathcal{R}) > 0$. Thus, in equilibrium, each firm will use all its data, i.e., $\mathcal{R}_i^{ml} =\mathcal{D}_i$. 

Substituting $\mathcal{R}_i^{ml} $  back into $p^{ml}_i(\mathcal{R}_i,\mathcal{R}_{-i})$ and $\Pi_i^{ml}(\mathcal{R}_i,\mathcal{R}_{-i})$ yields the optimal price and profit $p_i^{ml}$ and $\Pi_i^{ml}$ given in Lemma 1.$\hfill\qedsymbol$

\subsection{Proof of Lemma 2} 
The (expected) profit of firm $i$ is given by $\Pi_i^{fl}(p_i,p_{-i})=p_i\frac{v_i^{fl}(\mathcal{R}_i,\mathcal{R}_{-i})-p_i-\gamma  [v_{-i}^{fl}(\mathcal{R}_i,\mathcal{R}_{-i})-\!p_{-i}]}{1-\gamma ^2}$. By  solving $\frac{\partial \Pi_1^{fl}(p_1,p_2)}{\partial p_1}\!=\!0$ and $\frac{\partial \Pi_2^{fl}(p_1,p_2)}{\partial p_2}\!=\!0$,  we obtain firm $i$’ optimal prices (given the amount of data used $\mathcal{R}_1$ and $\mathcal{R}_2$) given by $p^{fl}_i(\mathcal{R}_i,\mathcal{R}_{-i})\!\!=\!\frac{  \left(2\!-\!\gamma ^2\right) v_i^{fl}(\mathcal{R}_i,\mathcal{R}_{-i})\!-\!\gamma  v_{-i}^{fl}(\mathcal{R}_i,\mathcal{R}_{-i})}{4-\gamma ^2} $. One can easily verify that the second-order condition holds and thus $p^{fl}_i(\mathcal{R}_i,\mathcal{R}_{-i})$ constitutes a local maximizer.

Substituting $p^{fl}_i(\mathcal{R}_i,\mathcal{R}_{-i})$ back to $\Pi_i^{fl}(p_i,p_{-i})$ yieldst $\Pi_i^{fl}(\mathcal{R}_i,\mathcal{R}_{-i})\!=\!\frac{\left[  \left(2\!-\!\gamma ^2\right) v_i^{fl}(\mathcal{R}_i,\mathcal{R}_{-i})\!-\!\gamma  v_{-i}^{fl}(\mathcal{R}_i,\mathcal{R}_{-i})\right]^2}{(1-\gamma^2)(4-\gamma ^2)^2}$. Firm $i$ chooses the optimal amount of data shared for training. One can easily verify that $\frac{\partial \Pi_i^{fl}(\mathcal{R}_i,\mathcal{R}_{-i})}{\partial \mathcal{R}_i}\!= \frac{2\left[(2-\gamma^2)v_i^{fl}(\mathcal{R}_i,\mathcal{R}_{-i})-\gamma v_{-i}^{fl}(\mathcal{R}_i,\mathcal{R}_{-i})\right]}{(1+\gamma)(2+\gamma)(2-\gamma)^2}\cdot \frac{\partial f(\mathcal{R}_i+\mathcal{R}_{-i})}{\partial \mathcal{R}_i}>0$ holds, because of (i) in Condition 1, $0\le \gamma\le 1$, and $f'(\mathcal{R}) > 0$. Thus, in equilibrium, each firm will share all its data, i.e., $\mathcal{R}_i^{fl} =\mathcal{D}_i$.

Substituting $\mathcal{R}_i^{fl} $  back into $p^{fl}_i(\mathcal{R}_i,\mathcal{R}_{-i})$ and $\Pi_i^{fl}(\mathcal{R}_i,\mathcal{R}_{-i})$ yields the optimal price and profit $p_i^{fl}$ and $\Pi_i^{fl}$ given in Lemma 2. $\hfill\qedsymbol$

\subsection{Proof of Lemma 3} 
From the results of Lemmas 1 and 2, one can easily verify that $\frac{\Pi_2^{fl}}{\Pi_2^{ml}}=[\frac{  \left(2\!-\!\gamma ^2\right) v_2^{fl}(\mathcal{D}_1,\mathcal{D}_{2})\!-\!\gamma  v_{1}^{fl}(\mathcal{D}_1,\mathcal{D}_{2})}{  \left(2\!-\!\gamma ^2\right) v_2^{ml}(\mathcal{D}_2)\!-\!\gamma  v_{1}^{ml}(\mathcal{D}_1)}]^2=\{1+\frac{  \left(2\!-\!\gamma ^2\right) [v_2^{fl}(\mathcal{D}_1,\mathcal{D}_{2})-v_2^{ml}(\mathcal{D}_2)]\!-\!\gamma[  v_{1}^{fl}(\mathcal{D}_1,\mathcal{D}_{2})-v_1^{ml}(\mathcal{D}_1)]}{  \left(2\!-\!\gamma ^2\right) v_2^{ml}(\mathcal{D}_2)\!-\!\gamma  v_{1}^{ml}(\mathcal{D}_1)}\}^2=\{1+\frac{  \left(2\!-\!\gamma ^2\right) [f(\mathcal{D}_1+\mathcal{D}_2)-f(\mathcal{D}_2)]\!-\!\gamma[f(\mathcal{D}_1+\mathcal{D}_2)-f(\mathcal{D}_1)]}{  \left(2\!-\!\gamma ^2\right) [v_2+f(\mathcal{D}_2)]\!-\!\gamma  [v_1+f(\mathcal{D}_1)]}\}^2>1$ holds, because of (ii) in Condition 1, $0 \le \gamma\le 1$, $\mathcal{D}_1>\mathcal{D}_2$ and $f'(\mathcal{D})>0$. However, it is inconclusive whether $\frac{\Pi_1^{fl}}{\Pi_1^{ml}} (=\{1+\frac{  \left(2\!-\!\gamma ^2\right) [f(\mathcal{D}_1+\mathcal{D}_2)-f(\mathcal{D}_1)]\!-\!\gamma[f(\mathcal{D}_1+\mathcal{D}_2)-f(\mathcal{D}_2)]}{  \left(2\!-\!\gamma ^2\right) [v_1+f(\mathcal{D}_1)]\!-\!\gamma  [v_2+f(\mathcal{D}_2)]}\}^2)$ is greater than 1, a result depending on the magnitudes of $\gamma$ and $\mathcal{D}_i$.
$\hfill\qedsymbol$

\subsection{Proof of Theorem 1} 
Solving $\frac{\Pi_1^{fl}}{\Pi_1^{ml}}>1$ and applying (ii) in Condition 1, $0 \le \gamma\le 1$, $\mathcal{D}_1>\mathcal{D}_2$ and $f'(\mathcal{\cdot})>0$, one can easily verify that condition (6) is needed to ensure that $\Pi_1^{fl}>\Pi_1^{ml}$ holds.
$\hfill\qedsymbol$

\subsection{Proof of Theorem  2} 

(i) The right-hand side of condition (6) is a decreasing function of \( \gamma \) and approaches $+\infty$ (1) as $\gamma$ approaches $0$ (1). Therefore, there exists a threshold degree of product substitution \( \gamma^* \in (0,1)\) such that $\frac{f(\mathcal{D}_1+\mathcal{D}_2)-f(\mathcal{D}_2)}{f(\mathcal{D}_1+\mathcal{D}_2)-f(\mathcal{D}_1)}=\frac{2-(\gamma^*)^2}{\gamma^*}$ and FL is formed if \( 0 < \gamma < \gamma^* \) (otherwise, ML is formed).

(ii) As $f'(\cdot) > 0$, $\frac{f(\mathcal{D}_1+\mathcal{D}_2)-f(\mathcal{D}_2)}{f(\mathcal{D}_1+\mathcal{D}_2)-f(\mathcal{D}_1)}$ ($=1+\frac{f(\mathcal{D}_1)-f(\mathcal{D}_2)}{f(\mathcal{D}_1+\mathcal{D}_2)-f(\mathcal{D}_1)}$) decreases in $\mathcal{D}_2$ and approaches $+\infty$ (1) when $\mathcal{D}_2$ approaches 0 $(\mathcal{D}_1)$. Therefore, there exists a threshold amount of data for firm 2 \( \mathcal{D}_2^* \in [0,\mathcal{D}_1)\) such that $\frac{f(\mathcal{D}_1+\mathcal{D}_2^*)-f(\mathcal{D}_2^*)}{f(\mathcal{D}_1+\mathcal{D}_2^*)-f(\mathcal{D}_1)}=\frac{2-\gamma^2}{\gamma}$ and FL is formed if  \( \mathcal{D}_2^* < \mathcal{D}_2< \mathcal{D}_1 \) (otherwise, ML is formed).

(iii) Since \( v_i \) is not included in condition (6), the (initial) product quality (\( v_i \)) does not affect firms' preference for ML or FL.
$\hfill\qedsymbol$

\subsection{Proof of Theorem  3} 

One can easily confirm that consumer surplus under ML and FL are given by Equation (\ref{cs}) and social welfare is given by
$SW^{ml}=CS^{ml}+\Pi_1^{ml}+\Pi_2^{ml}$ and $SW^{fl}=CS^{fl}+\Pi_1^{fl}+\Pi_2^{fl}$. 

\begin{equation}\label{cs}
\left\{
\begin{aligned}
CS^{ml}\!&=\!\frac{\left(4\!-\!3 \gamma ^2\right) \left[(v_1^{ml})^2\!+\!(v_2^{ml})^2\right]\!-\!2 \gamma ^3 v_1^{ml}v_2^{ml}}{2 \left(4-\gamma ^2\right)^2 \left(1-\gamma ^2\right)}\\
CS^{fl}\!&=\!\frac{\left(4\!-\!3 \gamma ^2\right) \left[(v_1^{fl})^2\!+\!(v_2^{fl})^2\right]\!-\!2 \gamma ^3 v_1^{fl}v_2^{fl}}{2 \left(4-\gamma ^2\right)^2 \left(1-\gamma ^2\right)}. \\
\end{aligned}
\right.
\end{equation}

Next, we prove that $CS^{fl}>CS^{ml}$. One can easily observe in (\ref{cs}) that $CS^{fl}$ and $CS^{ml}$ have the same functional form, except that $v_i^{fl}>v_i^{ml}$. Thus, we define a function $CS(v_1,v_2)\!=\!\frac{\left(4\!-\!3 \gamma ^2\right) \left(v_1^2\!+\!v_2^2\right)\!-\!2 \gamma ^3 v_1v_2}{2 \left(4-\gamma ^2\right)^2 \left(1-\gamma ^2\right)}$, and thus, proving $CS^{fl}>CS^{ml}$ is reduced to proving $\frac{\partial CS(v_1,v_2)}{\partial v_1}>0$ and $\frac{\partial CS(v_1,v_2)}{\partial v_2}>0$ hold at the same time. One can easily verify that $\frac{\partial CS(v_1,v_2)}{\partial v_1}=\frac{\left(4-3 \gamma ^2\right) v_1-\gamma ^3 v_2}{\left(4-\gamma ^2\right)^2 \left(1-\gamma ^2\right)}>0$, $\frac{\partial CS(v_1,v_2)}{\partial v_2}=\frac{\left(4-3 \gamma ^2\right) v_2-\gamma ^3 v_1}{\left(4-\gamma ^2\right)^2 \left(1-\gamma ^2\right)}>0$ hold because of $0 \!\le\! \gamma\!\le\! 1$ and (i) of Condition 1. Since $v_1^{fl}>v_1^{ml}$ and $v_2^{fl}>v_2^{ml}$, it follows that $CS(v_1^{fl},v_2^{fl})>CS(v_1^{ml},v_2^{ml})$, i.e., $CS^{fl}>CS^{ml}$ is proved. 

For social welfare, $SW^{ml}=CS^{ml}+\Pi_1^{ml}+\Pi_2^{ml}$, $SW^{fl}=CS^{fl}+\Pi_1^{fl}+\Pi_2^{fl}$. Because when FL is formed (condition (6) is satisfy) $\Pi_1^{fl}>\Pi_1^{ml}$ and $\Pi_2^{fl}>\Pi_2^{ml}$ and $CS^{fl}>CS^{ml}$, it follows that $SW^{fl}>SW^{ml}$. These results imply an “All-Win” situation in which all stakeholders (firms, consumers, and social planner) benefit from forming FL.\hfill\qedsymbol

\subsection{Proof of Theorem 4} 
From Lemma 3 and Theorem 2, one can easily confirm that, when $0<\gamma<\gamma^*$, $\Pi_2^{fl}>\Pi_2^{ml}$ and $\Pi_1^{fl}>\Pi_1^{ml}$, $\Delta_1+\Delta_2 \ge 0$ holds; when $\gamma=\gamma^*$, $\Pi_2^{fl}>\Pi_2^{ml}$ and $\Pi_1^{fl}=\Pi_1^{ml}$, $\Delta_1+\Delta_2 \ge 0$ holds too. However, when $\gamma^*<\gamma\le 1$, $\Delta_1+\Delta_2 \ge 0$ may not hold because $\Pi_2^{fl}>\Pi_2^{ml}$ but $\Pi_1^{fl}<\Pi_1^{ml}$. The following proof will demonstrate that there exists a threshold degree of product substitution ($\hat{\gamma} \in (\gamma^*,1)$) such that $\Delta_1+\Delta_2 \ge 0$ if \(\gamma^* < \gamma \le \hat{\gamma}\); otherwise, $\Delta_1+\Delta_2 < 0$. 

Specifically, proving $\Delta_1+\Delta_2<0$ is equivalent to proving the following two results (i) $\frac{\partial \left(\Delta_1+\Delta_2 \right)}{\partial \gamma}<0$ and (ii) $\Delta_1+\Delta_2 <0$ when $\gamma \rightarrow 1$, where $\Delta_1+\Delta_2=\frac{\left(4-3 \gamma ^2+\gamma ^4\right)t_1-4 \left(2 \gamma -\gamma ^3\right) t_2}{\left(1-\gamma ^2\right)\left(4-\gamma ^2\right)^2 }$, $t_1=(v_1^{fl})^2+(v_2^{fl})^2-(v_1^{ml})^2-(v_2^{ml})^2$ and $t_2=v_1^{fl} v_2^{fl}-v_1^{ml} v_2^{ml}$, and $\frac{\partial \left(\Delta_1+\Delta_2 \right)}{\partial \gamma}=\frac{2[(12\gamma-7\gamma^3+2\gamma^5-\gamma^7)t_1- 2 (8+2\gamma^2-7\gamma^4+3\gamma^6) t_2]}{\left(1-\gamma ^2\right)^2\left(4-\gamma ^2\right)^3 }$.

To establish result (i), i.e., $\frac{\partial \left(\Delta_1+\Delta_2 \right)}{\partial \gamma}<0$, one just need to prove $\frac{t_1}{2 t_2}<h(\gamma)$ holds, where $h(\gamma)= \frac{8+2\gamma^2-7\gamma^4+3\gamma^6}{12\gamma-7\gamma^3+2\gamma^5-\gamma^7}$. One can confirm that the left-hand side of the above inequality $\frac{t_1}{2 t_2}=\frac{(v_1^{fl})^2+(v_2^{fl})^2-(v_1^{ml})^2-(v_2^{ml})^2}{2(v_1^{fl} v_2^{fl}-v_1^{ml} v_2^{ml})}=1+\frac{(v_1^{fl}-v_2^{fl})^2-(v_1^{ml}-v_2^{ml})^2}{2(v_1^{fl} v_2^{fl}-v_1^{ml} v_2^{ml})}=1+\frac{(v_1-v_2)^2-[v_1+f(\mathcal{D}_1)-v_2-f(\mathcal{D}_2)]^2}{2(v_1^{fl} v_2^{fl}-v_1^{ml} v_2^{ml})}=1+\frac{[f(\mathcal{D}_2-f(\mathcal{D}_1)][2v_1+f(\mathcal{D}_1)-2v_2-f(\mathcal{D}_2)]}{2(v_1^{fl} v_2^{fl}-v_1^{ml} v_2^{ml})}<1$, because of $f'(\cdot)>0$ and $\mathcal{D}_1>\mathcal{D}_2$ while the right-hand side $\frac{\partial h(\gamma)}{\gamma}=\frac{3 \left(\gamma ^2-4\right)^2 \left(\gamma ^8+3 \gamma ^6-5 \gamma ^4+3 \gamma ^2-2\right)}{\gamma ^2 \left(\gamma ^6-2 \gamma ^4+7 \gamma ^2-12\right)^2}<0$ because of $0\le \gamma \le 1$. Therefore, $h(\gamma)\ge h(1)=1>\frac{t_1}{2 t_2}$. Result (i) is proved.

To prove result (ii), i.e., $\Delta_1+\Delta_2 <0$ when $\gamma \rightarrow 1$, one can easily confirm that, as $\gamma \rightarrow 1$, $\left(1-\gamma ^2\right)\left(4-\gamma ^2\right)^2 \rightarrow 0^+$ and $\left(4-3 \gamma ^2+\gamma ^4\right)t_1-4 \left(2 \gamma -\gamma ^3\right) t_2 \rightarrow 2(t_1-2t_2)<0$ because of $\frac{t_1}{2t_2}<1$. Therefore, $\Delta_1+\Delta_2=\frac{\left(4-3 \gamma ^2+\gamma ^4\right)t_1-4 \left(2 \gamma -\gamma ^3\right) t_2}{\left(1-\gamma ^2\right)\left(4-\gamma ^2\right)^2 } <0$ when $\gamma \rightarrow 1$. Result (ii) is proved.

Because both results (i) and (ii) hold, there must exist a threshold degree of product substitution ($\hat{\gamma} \in (\gamma^*,1)$) such that $\Delta_1+\Delta_2 \ge 0$ if \(\gamma^* < \gamma \le \hat{\gamma}\); otherwise, $\Delta_1+\Delta_2 < 0$. Thus, the results in Theorem 4 are proved.
$\hfill\qedsymbol$
\\ \hspace*{\fill} \\

\bibliography{aaai25}

\begin{thebibliography}{26}
\providecommand{\natexlab}[1]{#1}

\bibitem[{Bi, Gupta, and Yang(2023)}]{doi:10.1287/mnsc.2023.00611}
Bi, X.; Gupta, A.; and Yang, M. 2023.
\newblock Understanding Partnership Formation and Repeated Contributions in Federated Learning: An Analytical Investigation.
\newblock \emph{Management Science}.

\bibitem[{Blum et~al.(2021)Blum, Haghtalab, Phillips, and Shao}]{pmlr-v139-blum21a}
Blum, A.; Haghtalab, N.; Phillips, R.~L.; and Shao, H. 2021.
\newblock One for One, or All for All: Equilibria and Optimality of Collaboration in Federated Learning.
\newblock In Meila, M.; and Zhang, T., eds., \emph{Proceedings of the 38th International Conference on Machine Learning}, volume 139 of \emph{Proceedings of Machine Learning Research}, 1005--1014. PMLR.

\bibitem[{Bonawitz et~al.(2019)Bonawitz, Eichner, Grieskamp, Huba, Ingerman, Ivanov, Kiddon, Kone\v{c}n\'{y}, Mazzocchi, McMahan, Van~Overveldt, Petrou, Ramage, and Roselander}]{MLSYS2019_7b770da6}
Bonawitz, K.; Eichner, H.; Grieskamp, W.; Huba, D.; Ingerman, A.; Ivanov, V.; Kiddon, C.; Kone\v{c}n\'{y}, J.; Mazzocchi, S.; McMahan, B.; Van~Overveldt, T.; Petrou, D.; Ramage, D.; and Roselander, J. 2019.
\newblock Towards Federated Learning at Scale: System Design.
\newblock In Talwalkar, A.; Smith, V.; and Zaharia, M., eds., \emph{Proceedings of Machine Learning and Systems}, volume~1, 374--388.

\bibitem[{Chen et~al.(2024)Chen, Li, Liu, Zheng, Du, and Cheng}]{CHEN2024120527}
Chen, J.; Li, M.; Liu, T.; Zheng, H.; Du, H.; and Cheng, Y. 2024.
\newblock Rethinking the defense against free-rider attack from the perspective of model weight evolving frequency.
\newblock \emph{Information Sciences}, 668: 120527.

\bibitem[{Dixit(1979)}]{Dixit1979}
Dixit, A. 1979.
\newblock A Model of Duopoly Suggesting a Theory of Entry Barriers.
\newblock \emph{The Bell Journal of Economics}, 10(1): 20--32.

\bibitem[{Fraboni, Vidal, and Lorenzi(2021)}]{fraboni2021free}
Fraboni, Y.; Vidal, R.; and Lorenzi, M. 2021.
\newblock Free-rider attacks on model aggregation in federated learning.
\newblock In \emph{International Conference on Artificial Intelligence and Statistics}, 1846--1854. PMLR.

\bibitem[{Geyer, Klein, and Nabi(2018)}]{geyer2018differentiallyprivatefederatedlearning}
Geyer, R.~C.; Klein, T.; and Nabi, M. 2018.
\newblock Differentially Private Federated Learning: A Client Level Perspective.
\newblock arXiv:1712.07557.

\bibitem[{Kairouz et~al.(2021)Kairouz, McMahan, Avent, Bellet, Bennis, Bhagoji, Bonawitz, Charles, Cormode, and Cummings}]{kairouz2021}
Kairouz, P.; McMahan, H.~B.; Avent, B.; Bellet, A.; Bennis, M.; Bhagoji, A.~N.; Bonawitz, K.; Charles, Z.; Cormode, G.; and Cummings, R. 2021.
\newblock Advances and Open Problems in Federated Learning.
\newblock \emph{Foundations and Trends® in Machine Learning}, 14(1–2): 1--210.

\bibitem[{Karimireddy, Guo, and Jordan(2022)}]{karimireddy2022mechanismsincentivizedatasharing}
Karimireddy, S.~P.; Guo, W.; and Jordan, M.~I. 2022.
\newblock Mechanisms that Incentivize Data Sharing in Federated Learning.
\newblock arXiv:2207.04557.

\bibitem[{Konečný et~al.(2017)Konečný, McMahan, Yu, Richtárik, Suresh, and Bacon}]{konečný2017federatedlearningstrategiesimproving}
Konečný, J.; McMahan, H.~B.; Yu, F.~X.; Richtárik, P.; Suresh, A.~T.; and Bacon, D. 2017.
\newblock Federated Learning: Strategies for Improving Communication Efficiency.
\newblock arXiv:1610.05492.

\bibitem[{Li, Abbas, and Koutsoukos(2020)}]{NEURIPS2020_d37eb50d}
Li, J.; Abbas, W.; and Koutsoukos, X. 2020.
\newblock Byzantine Resilient Distributed Multi-Task Learning.
\newblock In Larochelle, H.; Ranzato, M.; Hadsell, R.; Balcan, M.; and Lin, H., eds., \emph{Advances in Neural Information Processing Systems}, volume~33, 18215--18225. Curran Associates, Inc.

\bibitem[{Lyu et~al.(2020{\natexlab{a}})Lyu, Xu, Wang, and Yu}]{Lyu2020}
Lyu, L.; Xu, X.; Wang, Q.; and Yu, H. 2020{\natexlab{a}}.
\newblock \emph{Collaborative Fairness in Federated Learning}, 189--204.
\newblock Cham: Springer International Publishing.
\newblock ISBN 978-3-030-63076-8.

\bibitem[{Lyu et~al.(2020{\natexlab{b}})Lyu, Yu, Nandakumar, Li, Ma, Jin, Yu, and Ng}]{9098045}
Lyu, L.; Yu, J.; Nandakumar, K.; Li, Y.; Ma, X.; Jin, J.; Yu, H.; and Ng, K.~S. 2020{\natexlab{b}}.
\newblock Towards Fair and Privacy-Preserving Federated Deep Models.
\newblock \emph{IEEE Transactions on Parallel and Distributed Systems}, 31(11): 2524--2541.

\bibitem[{McMahan et~al.(2017)McMahan, Moore, Ramage, Hampson, and Arcas}]{pmlr-v54-mcmahan17a}
McMahan, B.; Moore, E.; Ramage, D.; Hampson, S.; and Arcas, B. A.~y. 2017.
\newblock {Communication-Efficient Learning of Deep Networks from Decentralized Data}.
\newblock In Singh, A.; and Zhu, J., eds., \emph{Proceedings of the 20th International Conference on Artificial Intelligence and Statistics}, volume~54 of \emph{Proceedings of Machine Learning Research}, 1273--1282. PMLR.

\bibitem[{Richardson, Filos-Ratsikas, and Faltings(2020)}]{Richardson2020}
Richardson, A.; Filos-Ratsikas, A.; and Faltings, B. 2020.
\newblock \emph{Budget-Bounded Incentives for Federated Learning}, 176--188.
\newblock Cham: Springer International Publishing.
\newblock ISBN 978-3-030-63076-8.

\bibitem[{Sheller et~al.(2020)Sheller, Edwards, Reina, Martin, Pati, Kotrotsou, Milchenko, Xu, Marcus, Colen, and Bakas}]{Sheller2020FederatedLI}
Sheller, M.~J.; Edwards, B.; Reina, G.~A.; Martin, J.; Pati, S.; Kotrotsou, A.; Milchenko, M.; Xu, W.; Marcus, D.; Colen, R.~R.; and Bakas, S. 2020.
\newblock Federated learning in medicine: facilitating multi-institutional collaborations without sharing patient data.
\newblock \emph{Scientific Reports}, 10.

\bibitem[{Singh and Vives(1984)}]{Vives1984}
Singh, N.; and Vives, X. 1984.
\newblock Price and Quantity Competition in a Differentiated Duopoly.
\newblock \emph{The RAND Journal of Economics}, 15(4): 546--554.

\bibitem[{Tan et~al.(2023)Tan, Yu, Cui, and Yang}]{9743558}
Tan, A.~Z.; Yu, H.; Cui, L.; and Yang, Q. 2023.
\newblock Towards Personalized Federated Learning.
\newblock \emph{IEEE Transactions on Neural Networks and Learning Systems}, 34(12): 9587--9603.

\bibitem[{Wang et~al.(2022)Wang, Chang, Rodrìguez, and Wang}]{9912903}
Wang, J.; Chang, X.; Rodrìguez, R.~J.; and Wang, Y. 2022.
\newblock Assessing Anonymous and Selfish Free-rider Attacks in Federated Learning.
\newblock In \emph{2022 IEEE Symposium on Computers and Communications (ISCC)}, 1--6.

\bibitem[{Wu and Yu(2022)}]{9809786}
Wu, X.; and Yu, H. 2022.
\newblock MarS-FL: Enabling Competitors to Collaborate in Federated Learning.
\newblock \emph{IEEE Transactions on Big Data}, 1--11.

\bibitem[{Yang et~al.(2019)Yang, Liu, Chen, and Tong}]{Yang2019}
Yang, Q.; Liu, Y.; Chen, T.; and Tong, Y. 2019.
\newblock Federated Machine Learning: Concept and Applications.
\newblock \emph{ACM Trans. Intell. Syst. Technol.}, 10(2).

\bibitem[{Zeng et~al.(2021)Zeng, Zeng, Wang, Li, and Chu}]{Zeng2021ACS}
Zeng, R.; Zeng, C.; Wang, X.; Li, B.; and Chu, X. 2021.
\newblock A Comprehensive Survey of Incentive Mechanism for Federated Learning.
\newblock \emph{ArXiv}, abs/2106.15406.

\bibitem[{Zhan et~al.(2021)Zhan, Li, Guo, and Qu}]{9409833}
Zhan, Y.; Li, P.; Guo, S.; and Qu, Z. 2021.
\newblock Incentive Mechanism Design for Federated Learning: Challenges and Opportunities.
\newblock \emph{IEEE Network}, 35(4): 310--317.

\bibitem[{Zhang, Ma, and Chen(2023)}]{9705098}
Zhang, N.; Ma, Q.; and Chen, X. 2023.
\newblock Enabling Long-Term Cooperation in Cross-Silo Federated Learning: A Repeated Game Perspective.
\newblock \emph{IEEE Transactions on Mobile Computing}, 22(7): 3910--3924.

\bibitem[{Zhang and Huang(2024)}]{zhang2024adversarial}
Zhang, R.-X.; and Huang, T. 2024.
\newblock Adversarial Attacks on Federated-Learned Adaptive Bitrate Algorithms.
\newblock In \emph{Proceedings of the AAAI Conference on Artificial Intelligence}, volume 38(1), 419--427.

\bibitem[{Zhu et~al.(2021)Zhu, Shu, Zou, and Jia}]{zhu2021advanced}
Zhu, Z.; Shu, J.; Zou, X.; and Jia, X. 2021.
\newblock Advanced free-rider attacks in federated learning.
\newblock In \emph{the 1st NeurIPS Workshop on New Frontiers in Federated Learning Privacy, Fairness, Robustness, Personalization and Data Ownership}.

\end{thebibliography}

\end{document}